%% file: main.tex
\documentclass[runningheads]{llncs}

\input{dependencies}
\input{macros-small}

\input{safe_macros}

\input{macros-file-formats}

\title{Computing the Death Rate of COVID-19}
\author{
  Naveen Pai\orcidID{0000-0002-5980-560X}
  \and 
  Sean Zhang\orcidID{0000-0003-1699-1809}
  \and
  Mor Harchol-Balter\orcidID{0000-0003-1721-6759}
  \thanks{This research was supported by NSF-CMMI-1938909, NSF-CSR-1763701, and a Google 2020 Faculty Research Award.} 
}
\institute{
\email{\{nvpai, xiaoronz, harchol\}@andrew.cmu.edu}
\\
Carnegie Mellon University
}
\date{}

\begin{document}

\maketitle

\section*{Abstract}

The Infection Fatality Rate (IFR) of COVID-19 is difficult to estimate because the number of infections is unknown and there is a lag between each infection and the potentially subsequent death. We introduce a new approach for estimating the IFR by first estimating the entire sequence of daily infections. Unlike prior approaches, we incorporate existing data on the number of daily COVID-19 tests into our estimation; knowing the test rates helps us estimate the ratio between the number of cases and the number of infections. Also unlike prior approaches, rather than determining a constant lag from studying a group of patients, we treat the lag as a random variable, whose parameters we determine empirically by fitting our infections sequence to the sequence of deaths.   Our approach allows us to narrow our estimation to smaller time intervals in order to observe how the IFR changes over time. We analyze a 250 day period starting on March 1, 2020.
 We estimate that the IFR in the U.S. decreases from a high of $0.68\%$ down to $0.24\%$ over the course of this time period. We also provide IFR and lag estimates for Italy, Denmark, and the Netherlands, all of which also exhibit decreasing IFRs but to different degrees.

\section{Introduction} \label{intro}

The COVID-19 pandemic has raged for over a year now, greatly disrupting the lives of people all over the world.   While much research has been done on modeling the spread of the disease,  one particular question that remains unresolved is:  
{\em What is the death rate of COVID-19? }

A good estimate for the death rate could better inform government policies, but it has been surprisingly difficult to estimate the death rate \cite{why-important}. 
When we talk about the ``death rate" we will be talking about the probability that an infected person ends up dying.    This is formally called the Infection Fatality Rate (IFR) and will be defined in Section \ref{sec:model}.

\subsection{Challenges in estimating the death rate}\label{sec:introchallenges}

Several challenges arise in estimating the death rate.  These include:

{\bf Cases are recorded, not infections:} Estimating the death rate requires knowing the number of deaths relative to the number of infections.   Unfortunately, what is reported is the number of daily cases, not the number of daily infections.  Note that a {\em case} is a test that results in a positive, however not all infected people are necessarily tested.  Thus, determining the death rate firstly involves creating some {\em estimate} of the number of infections.  One might think that testing can provide such an estimate, but testing is not randomized \cite{not-randomized}, so the proportion of test-takers who are infected is not indicative of the infection rate in the overall population.

{\bf Lag between cases and deaths:}  In this paper we refer to the time between when a person tests positive (a {\em case}) and the potentially subsequent death as the {\em lag time}\footnote{In some literature, {\em lag} is described as the time from initial infection to death, but such studies involve only a small group of patients, for whom the initial infection time is approximately known.  In general it is impossible to ascertain the time of the initial infection from large-scale reports.}.  Unfortunately, this lag time varies for each individual. To make matters worse, lag time is also affected by larger trends, such as the age group that is most commonly infected and the treatments that are available \cite{who-report}.  Thus, the mean lag time may change as the pandemic progresses.

{\bf The testing rate is not constant:} The number of daily cases can provide insight into the number of infections.   However, the number of cases also depends on the rate at which people get tested for COVID.  Higher testing rates lead to more cases being observed. The fact that the testing rate has varied greatly over time \cite{testing-change} makes it difficult to interpret the number of cases.

{\bf The death rate varies over time:}  The death rate is itself not constant over time.  The death rate goes down when hospitals find new treatments, but then it shoots back up when hospitals get overloaded and run out of workers, oxygen, or ventilators \cite{death-rate-change}. The death rate grows when older people in nursing homes are being most commonly infected, but shrinks when younger people are most commonly infected \cite{death-rate-change}.   Because the death rate varies, simply trying to fit one number to all the data is not necessarily the best approach.

{\bf Reported numbers are not always accurate:} Finally, we are always dealing with reported numbers, not true numbers.   It is well known that in many states there is a delay in recording deaths and reporting them \cite{states-slow}.  Some COVID cases never get reported at all. For example, a person may die from COVID without ever being tested     \cite{deaths-unattributed}.  Finally, there are both false positives and false negatives in the reported cases \cite{false-pos}.

{\bf Antibodies can wear off:} People who were infected and recovered may lose their antibodies after a few months \cite{antibodies-fade}. This makes it difficult to use antibody studies to estimate how many people have been infected by COVID.

\subsection{Prior approaches to estimating the death rate}

There has been some prior work on determining the death rate, either within the U.S., or in other countries (see Section~\ref{sec:prior} for details). Much of the prior work is not actually looking at the Infection Fatality Rate (IFR), but rather at the Case Fatality Rate (CFR), which is the fraction of positive cases that result in deaths. Our goal in this paper is to determine the IFR.    

For those works that do directly try to measure the IFR, most follow this simplistic 2-step approach which is not time-dependent:
\begin{enumerate}
    \item Estimate the total number of infections by time $t$ by looking at the results of antibody tests at time $t$. 
    \item Divide the total number of deaths by time $t$ (this is easily found in reports) by the total number of infections by time $t$ from the previous step.   This yields the estimated death rate during the period $[0,t]$.
\end{enumerate}
Sometimes the authors go a step further, by incorporating a lag between the case and death (or the infection and death).  But this lag is often assumed to be a {\em fixed} constant value, estimated via small {\em in-person} studies.

\subsection{Drawbacks of the prior approaches}

There are several drawbacks to prior approaches.
Firstly, the prior approaches are people-intensive, which is costly. 
Secondly, the prior approaches are limited in what they can produce.   The IFR can only be estimated on days when an antibody study was conducted. To make matters worse, antibody studies become less accurate over time, since people lose antibodies, causing an underestimate of past infections \cite{antibodies-fade}. 
But the biggest drawback of the prior work is that it doesn't take into account the changes in the pandemic.  As we've explained, the lag time changes over time.   The IFR itself changes over time.  
Prior approaches are not well-suited to take these temporal changes into account.

\subsection{A new data-driven approach to estimating the death rate}

In this paper we present a data-driven approach.  Unlike prior approaches, our method does not rely on having a set of patients whom we can track and study.  Instead, our approach is solely based on studying readily available numbers on \emph{Our World in Data} \cite{owid} along with a single antibody study that can be conducted at an arbitrary time. By relying on a single antibody study, we can ignore later antibody studies which underestimate infections due to antibodies fading. Secondly, our approach easily admits temporal changes; in particular, we incorporate changes in the lag distribution, changes in the testing rate, and a time-varying IFR.  
The data that we use includes the sequence of daily deaths, daily cases, and daily tests, over a period of 250 days beginning with March 1, 2020, as well as antibody results that were available two or three months after March 1, 2020. 

Our first key idea is that we need to estimate the entire sequence of daily infections.  The sequence itself is needed because it allows us to see how the IFR changes over time and to study the temporal changes in the lag time distribution.  

Our second key idea is that the sequence of daily infections can be approximated using a function of both the daily cases and tests. This function is increasing with respect to the number of cases but decreasing with respect to the number of tests.  It furthermore assumes that a person who is infected is more likely to get tested than a person who is not infected. Section \ref{sec:inf-est} explains the intuition behind our function.  We find that the appropriate parameters of the function can be determined empirically via antibody results.

Once we have our estimated infection sequence, our next key idea is that we should determine the lag time {\em and} the IFR {\em concurrently}, not individually.   Every possible choice of lag time can be viewed as a time shift of the infection sequence, while every possible choice of the IFR can be viewed as a vertical scaling of the infection sequence.  Together, every choice of lag time and IFR results in a particular ``candidate" death sequence.    To find the ``best" choice of lag time and IFR, we simply pick the parameters which result in a candidate death sequence which best fits the given death sequence in our data.  We actually take this idea a step further and allow the lag time to follow a {\em distribution} whose parameters are incorporated in the above optimization process. 

Our final idea is that this entire optimization process is best done by segmenting time into intervals (we use intervals of 50 days), because we find that the IFR and the lag time change over time.  
We repeat the above approach over several different countries, starting with the U.S. and moving on to Italy, Denmark, and the Netherlands.  There is no reason why our approach can't be applied to other countries as well, provided that the data is available.  

\subsection{Synopsis of our Findings}

Figure \ref{fig:intropredictionsUSA} illustrates a synopsis of our findings for the United States.
In Figure \ref{fig:intropredictionsUSA}(a), we see the sequence of daily cases (in blue), and above it we see our estimated daily infection sequence (in purple).   Importantly, our estimated daily infection sequence is {\em not} a mere vertical scaling.   The reason why is that in deriving our infection sequence we incorporate the time-varying testing rate, not shown here (see Section \ref{sec:evaluation}).

\evaluatedcountrypredictions{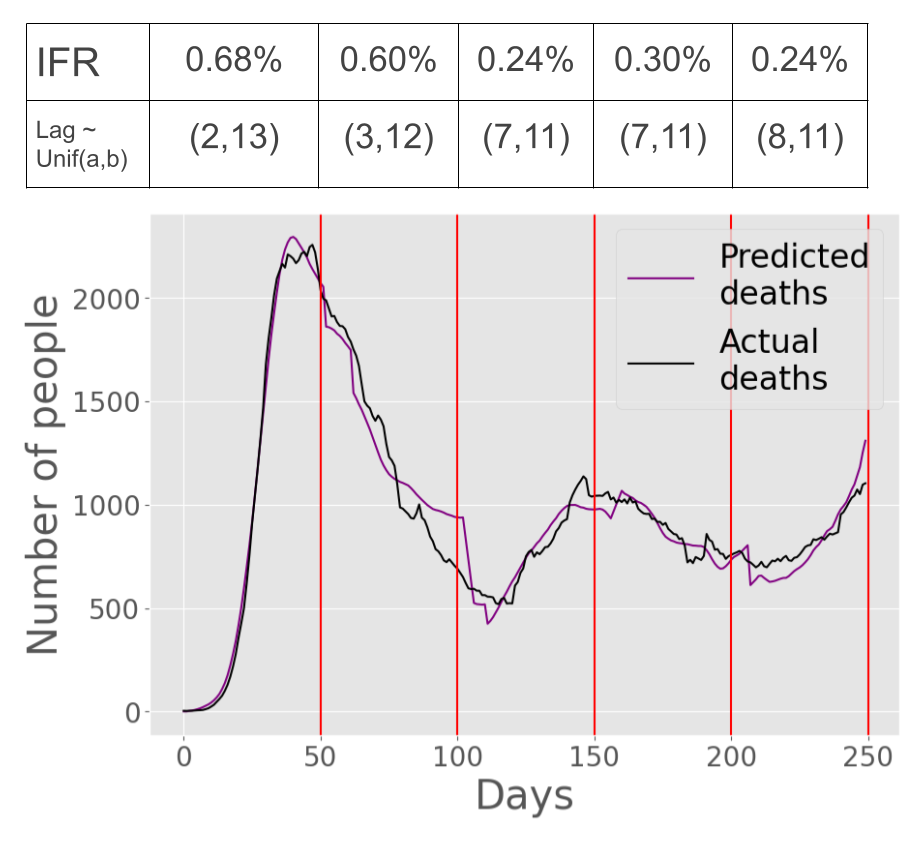}{fig:intropredictionsUSA}{From cases to estimated infections to predicted IFR and lag in the U.S.}

In Figure \ref{fig:intropredictionsUSA}(b), we illustrate our best ``candidate" death sequence (which we derive from our estimated infection sequence as described in Section \ref{sec:inf-est}), and we compare it with the true death sequence.   These two sequences are a good fit, indicating that our estimated IFR and lag are accurate. We annotate the top of Figure \ref{fig:intropredictionsUSA}(b) with our estimated IFR and the lag distribution in days for each 50-day interval. We find that the IFR decreases over time, eventually changing from 0.68\% to 0.24\%. We find that the lag is well fit by a $\Uniform(a,b)$ distribution, with a mean lag of about 8 days throughout the pandemic. Section \ref{sec:evaluation} contains more details and also repeats this process for Italy, Denmark, and the Netherlands.  


\subsection{Road map for the rest of the paper}

Section \ref{sec:prior} discusses the prior work in more detail. In Section \ref{sec:model} we provide definitions and notation. Section \ref{sec:inf-est} describes how we estimate a sequence of daily infections based on reported sequences for daily cases, tests, and deaths. In Section \ref{sec:infer-ifr-lag} we present our algorithm for calculating the IFR and lag time, and in Section \ref{sec:smaller-intervals} we refine our algorithm to look at smaller time intervals. In Section \ref{sec:evaluation}, we apply our approach to various countries. 
We conclude in Section \ref{sec:conclusion} and discuss opportunities for future work.

\section{Prior work and how we differ}\label{sec:prior}

{\bf Most prior work estimates CFR}, the case fatality rate, rather than IFR, which is the focus of this paper. Several studies estimate the CFR while adjusting for lag between cases and deaths. Newall et al. \cite{Newall} estimates the lag-adjusted CFR using a lognormal distribution for the lag with mean 14.5 days, which was estimated using the onset to death time of 34 patients, based on the case data from the Diamond Princess cruise ship.
Using a very similar approach, Russell et al. \cite{Russell} estimates the lag-adjusted CFR in South Korea. They used the first 66 deaths reported in South Korea to fit a lognormal distribution for the lag, which is then used to compute the adjusted CFR. 

By contrast our approach does not need the data of individual patients: we estimate the lag distribution using the sequence of cases and deaths only, which allows us to make use of a larger amount of data from different regions. We find that assuming lag to be uniformly distributed yields a slightly better fit in our model than assuming a lognormal distribution.   

{\bf Prior work on estimating IFR:}
Regarding IFR, various studies attempt to estimate this value using the data from antibody studies, in which a group of patients is tested for antibodies in an attempt to estimate the fraction of the total population that has been infected. 
The IFR is then estimated as the ratio between the fraction of the population that died and the fraction that was infected.   

However, {\em unlike our work, these studies do not incorporate knowledge of the time-varying testing rate}.    By not taking into account the testing rate, they end up with very different results than ours.
Meyerowitz-Katz et al.\ \cite{MeyerowitzKatz}
gives a comprehensive review of such studies which estimate the IFR.   
In particular, one study by Villa et al.\ \cite{Villa} estimates an IFR of 1.1\% for Italy up to the end of March 2020. By contrast, our estimated IFR for Italy is 2.2\% around March 2020. 

Several studies, e.g., Levin et al. \cite{Levin} and Marra and Quartin \cite{Marra}, take lag into account in determining the IFR, but they assume a {\em fixed lag}.   
By contrast, our work assumes a {\em lag distribution} with parameters that can vary by region and by time. Our more flexible approach accounts for differences in the spread and reporting of the disease. 

\section{Definitions and Notation}\label{sec:model}

Throughout this document, the terms {\em death rate} and {\em infection fatality rate (IFR)} are synonymous and are defined as follows:
Consider a period of time $t \in [a, b]$, where $a$ and $b$ denote particular days.  We consider the infections that occur during $[a,b]$ and the subsequent deaths, possibly after day $b$, that are a consequence of those infections.   We define:
\[
\text{IFR}[a,b] = \frac{\text{\# of people infected during  } [a,b] \text{ who died}}{\text{\# of infections during the interval } [a,b]} \ .
\]

In this document, we will make use of publicly-available data on the number of daily cases and the number of daily deaths in a range of countries \cite{data-JHU} (recall that a {\em case} denotes a positive test result). 
We let $\vec{c} = (c_1, \dots, c_k)$ be the time series of new cases per day for each of the first $k$ days. For our data, day 1 corresponds to March 1, 2020, and day $k=250$ corresponds to November 6, 2020. Similarly  $\vec{d} = (d_1, \dots, d_k)$ denotes the number of new deaths per day from COVID for each of the first $k$ days. We will also utilize reports of the number of daily tests \cite{testing-change}, which we denote by $\vec{t} = (t_1, \dots, t_k)$. 
We use $\vec{{i}} = ({i}_1, \dots, {i}_k)$ to denote our \emph{estimated} number of new infections per day.  




\section{Estimating the infections sequence}\label{sec:inf-est}

In this section, we explain how we determine our estimated sequence of infections: 
$\vec{{i}} = ({i}_1, \dots, {i}_k)$. 
At a high level, we start with the reported number of cases each day: $\vec{c} = (c_1, \dots, c_k)$. We then incorporate the number of tests each day, $\vec{t} = (t_1, \ldots, t_k)$ and also the results of a one-time antibody test.
These three pieces of information give us everything we need.


We will make one simplifying assumption that is needed only for analytical convenience but does not affect the derivation of IFR.   Our assumption is that if a person is  infected on day $j$, they either get tested on day $j$, or never get tested at all.  In reality, they might be tested any time after day $j$, but it will be convenient for us to assume that the testing happens on day $j$.  Note that this assumption will mean that our infections sequence will be slightly shifted forward in time from the true infections sequence.  However this will not affect our IFR because, in fitting the infections sequence to the deaths sequence, we still have a ``lag" variable that lets us account for an arbitrary time from case to death.  

With this simplifying assumption in mind, let us consider an arbitrary day $j$. Now let's assume we pick a random person on day $j$.  Let $I_j$ be the event that this randomly chosen individual becomes infected on day $j$. So $\p{I_j}=\frac{i_j}{N}$ where $N$ is the population of the country in consideration.  Now let $T_j$ be the event that this same randomly chosen individual is tested on day $j$.  So $\p{T_j}=\frac{t_j}{N}$. Conditional probability tells us that
\begin{equation*}
    \p{I_j}\cdot\p{T_j | I_j}=\p{T_j \cap I_j} \ . 
\end{equation*}
Notice that $T_j \cap I_j$ is the event that the randomly chosen individual is both infected and tested on day $j$ -- meaning that the individual becomes a ``case" on day $j$ (we assume that the test is always accurate).  So $\p{T_j \cap I_j}=\frac{c_j}{N}$. Putting this all together, we get that
\begin{equation}\label{eq:intermediateijformula}
    \frac{i_j}{N}\cdot\p{T_j | I_j} = \frac{c_j}{N} \ . 
\end{equation}
Thus we can approximate $i_j=\frac{c_j}{\p{T_j | I_j}}$ if we know  $\p{T_j | I_j}$.  

Now $\p{T_j | I_j}$ represents the fraction of people who are tested, given that they were infected on day $j$.  We will approximate $\p{T_j | I_j}$ as being a function of $\p{T_j}$, since $\p{T_j}=\frac{t_j}{N}$ is a known quantity which should be closely related to $\p{T_j | I_j}$.   To understand the relationship between $\p{T_j | I_j}$ and $\p{T_j}$, we first note that infected individuals are more likely to get tested than randomly chosen individuals, since infected individuals may be prompted to get tested by symptoms or contact-tracing. So $\p{T_j | I_j} > \p{T_j}$. Now consider the ratio between $\p{T_j | I_j}$ and $\p{T_j}$.  This ratio should be highest ($\gg 1$) when $\p{T_j}$ is low, because when testing is scarce only people with symptoms will be tested.   This ratio should be lowest (converging to $1$) when $\p{T_j}$ gets high, because at that point, everyone is being tested, regardless of whether they're infected or not.

\begin{figure}
    \centering
    \includegraphics[scale=0.3]{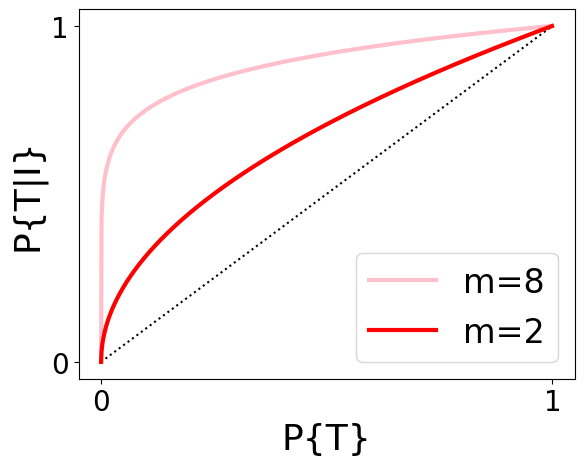}
    \caption{Red and pink solid curves show our assumed relationship between $\p{T_j | I_j}$ and $\p{T_j}$, from Equation \ref{eqn:1/m}, for a few values of $m$.}
    \label{fig:prob_relationship}
\end{figure}

Figure~\ref{fig:prob_relationship} illustrates a family of possible relationship between $\p{T_j | I_j}$ and $\p{T_j}$ that satisfies all these intuitions: 
\begin{eqnarray}
\p{T_j | I_j} &= & (\p{T_j})^{1/m} \ . \label{eqn:1/m}
\end{eqnarray}
We will assume this functional relationship holds for some parameter $m$, and determine the appropriate $m$ empirically. We tried many possible functions, and found that the family of curves in Equation~\ref{eqn:1/m} yielded the best fit in all of the countries we tried. Thus, via Equation \ref{eq:intermediateijformula} and Equation~\ref{eqn:1/m}, we conclude the following relationship between $i_j$, $c_j$, and $t_j$:
\begin{equation}\label{eq:calculateij}
    i_j = \frac{c_j}{\left(\frac{t_j}{N}\right)^{1/m}} \ . 
\end{equation}

We will now propose a method for empirically inferring the correct value of $m > 1$. An antibody study tells us, for some day $\ell$, the number of people $A$ who were infected prior to day $\ell$. Our estimated infections sequence should agree with this figure. Thus we can search for a value of $m$ for which:
\begin{equation}
    \sum_{j=1}^\ell i_j = \sum_{j=1}^\ell \frac{c_j}{\left(\frac{t_j}{N}\right)^{1/m}} \approx A \ . \label{eqn:sumijs}
\end{equation}
Notice that $\sum_{j=1}^\ell \frac{c_j}{\left(\frac{t_j}{N}\right)^{1/m}}$ is decreasing with respect to $m$. So, in practice, to find the appropriate value of $m$, we can do a binary search for $m$,  repeatedly evaluating  $\sum_{j=1}^\ell \frac{c_j}{\left(\frac{t_j}{N}\right)^{1/m}}$ until Equation~\ref{eqn:sumijs} is satisfied.

\section{Inferring IFR and lag using the infections sequence and deaths sequence}\label{sec:infer-ifr-lag}

At this point, we know our estimated sequence of infections $\vec{i} = ({i}_1, \dots, {i}_k)$. In this section, we describe our approach for estimating lag and IFR concurrently. 

Our high-level approach is as follows: 
We assume that the lag time between infections and deaths is a random variable $L$  following a $\Uniform(a,b)$ distribution with unknown parameters, $a$ and $b$, that we will determine.\footnote{We also considered other distributions for lag, such as the LogNormal($\mu$, $\sigma^2$) and Binomial($n$, $p$), but the best results were achieved with the $\Uniform(a,b)$. }
We start with the infections sequence $\vec{i}$. Consider shifting $\vec{i}$ by a particular lag distribution, $L$, and then vertically scaling the shifted sequence by a particular IFR\footnote{Although the lag, $L$, is being applied to the infections sequence $\vec{i}$, the simplifying assumption that we made in Section~\ref{sec:inf-est} while computing $\vec{i}$  implies that we should think of $L$ as representing a lag between {\em cases} and deaths.}.   The shifted and then scaled sequence is what we call a \emph{candidate death sequence}.  We want to choose the candidate death sequence (with its corresponding IFR and $L$) that best fits the actual death sequence, $\vec{d} = (d_1, \dots, d_k)$.  
We return the ``best-fit" $L$ and IFR as our final estimate.   
Part of our process will involve iterating over all possible values for $(a,b)$, assuming some generous upper bound on the lag.

We now explain our approach in more detail.
We first define a time shift, $\shift{L}{i}$, algorithmically. For each infection in $\vec{i}$, we will sample a new instance $\l$ from $L$ and shift the infection forward by $\l$ days. This produces a sequence $\shift{L}{i} = (I_1, I_2, I_3, \dots, I_{k})$ where each entry is a random variable denoting the number of shifted infections that fall on that day. 

Before we can compare the sequence of random variables, $\shift{L}{i}$, to the actual death sequence $\vec{d}$, we need to turn $\shift{L}{i}$ into a deterministic sequence, $\shiftexp{L}{i}$. We define the \textit{unscaled candidate death sequence}, $\shiftexp{L}{i}$ as
\begin{equation}\label{eq:defn_shift_rv}
\shiftexp{L}{i}\equiv(\E{I_1}, \E{I_2},
\E{I_3},\dots,\E{I_k}) \equiv (i'_1, i'_2, i'_3, \ldots, i'_k).
\end{equation}
To calculate each element $i'_j$ in $\shiftexp{L}{i}$, we apply linearity of expectations via:
\begin{equation}\label{eq:shift_rv_exp}
i'_j \equiv \E{I_j} = \sum_{w=1}^{j} i_w \cdot \p{L = j-w}.
\end{equation}
Once we have the deterministic sequence $\shiftexp{L}{i}$, we will want to compare it to our actual death sequence $\vec{d}$. We define an error metric by
$d(\vec{x}, \vec{y}) \equiv \sum_j (x_j - y_j)^2$.
Given a particular lag distribution $L$, we now estimate the IFR to be the optimal parameter $r$ which minimizes the error between $r\cdot\shiftexp{L}{i}$ and $\vec{d}$:
\[
\text{IFR}_L
\equiv \argmin_r d\left(r \cdot \shiftexp{L}{i}, \vec{d}\right)
= \argmin_r \sum_{j=1}^k (r \cdot i'_j  - d_j)^2.
\]
The right-hand side is quadratic in $r$ and has a unique minimum achieved at
\begin{equation}
\label{eq:closedformdist}
\text{IFR}_L=\frac{i'_1 d_1 + \cdots + i'_k d_k}{{i'_1}^2 + \cdots + {i'_k}^2}
= \frac{\shiftexp{L}{i} \cdot \vec{d}}{\norm{\shiftexp{L}{i}}^2}.
\end{equation}
To find the best candidate death sequence overall, we loop through all choices of the lag distribution, $L \sim \Uniform(a,b)$. We assume any reasonable lag to be well under 50 days, so we restrict $a\le b\le 50$. For each pair $(a,b)$, we compute the error given by $d(\text{IFR}_L \cdot \shiftexp{L}{i}, \vec{d})$. We choose the candidate death sequence with the smallest error, and output the IFR and $L$ corresponding to this best candidate death sequence. 
We describe the entire procedure of estimating $L$ and IFR in Algorithm \ref{alg:rv_lag}. 

\begin{algorithm}[H]
\caption{$\bestfit(\vec{i},\vec{d})$}
\label{alg:rv_lag}
\begin{algorithmic}[1]
\STATE $M^* \leftarrow \infty$, $a^* \leftarrow \infty$, $b^* \leftarrow \infty$, $r^* \leftarrow \infty$
\FOR{$a$ in $\{0, \dots, 50\}$}
\FOR{$b$ in $\{a, \dots, 50\}$}
    \STATE Assume $L\sim$ Uniform$(a,b)$
    \STATE Compute $\shiftexp{L}{i}$ via Equations \ref{eq:defn_shift_rv} and \ref{eq:shift_rv_exp}
    \label{bestfitshift}
    \STATE $r \leftarrow \frac{\shiftexp{L}{i}\cdot \vec{d}}{\|\shiftexp{L}{i}\|^2}$ \COMMENT{See Equation \ref{eq:closedformdist}}
    \STATE $M \leftarrow d(r\cdot \shiftexp{L}{i},\vec{d})$ 
    \IF{$M < M^*$}
        \STATE $M^* \leftarrow M$, $a^* \leftarrow a$, $b^* \leftarrow b$, $r^* \leftarrow r$
    \ENDIF
\ENDFOR    
\ENDFOR
\RETURN{$a^*$, $b^*$, $r^*$}
\end{algorithmic}
\end{algorithm}


\section{Inferring IFR in smaller time intervals}
\label{sec:smaller-intervals}
Our algorithm thus far assumes that the IFR and lag remain constant throughout the pandemic. In reality, the IFR and lag may change during the pandemic, due to a variety of conditions \cite{death-rate-change,who-report}. We will now outline an approach for estimating a time-varying IFR and lag.  Throughout our experiments we limit our results to $k = 250$ days, because that is the extent of the data that was available to us at the time of writing this paper.

We refer to our original IFR as IFR$[1,k]$.
In this section our goal is to find the best IFR and lag for 
smaller intervals of length $w < k$, where we imagine that $k$ is a multiple of $w$. In practice, we will let $w=50$, because we find that this interval size is small enough to accurately detect changes in IFR without overfitting.  We will derive IFR$[1,w]$, IFR$[w+1, 2w]$, $\ldots$, IFR$[k-w+1,k]$. 

\begin{figure}[h]
\centering
\includegraphics[scale=0.35]{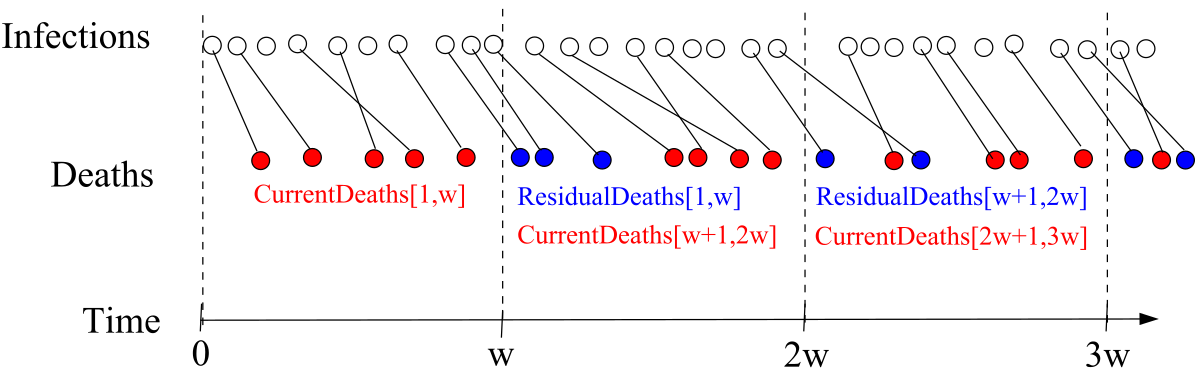}
\caption{Time is broken into intervals of length $w$.  For each interval, we show the current deaths in red and the residual deaths from the prior interval in blue.  Note that some infections do not result in deaths.   Note also that, because the lag is a random variable, the residual deaths and current deaths are sometimes interleaved.}
\label{fig:intervals}
\end{figure}

Our basic idea is to apply \bestfit (Algorithm \ref{alg:rv_lag}) to intervals of length $w$, while accounting for the fact that some deaths may be attributed to infections from the {\em prior} interval.
We call \emph{current deaths of $[1,w]$} the sequence of deaths which occurred in interval $[1,w]$ that are attributed to infections in $[1,w]$. We define the \emph{residual deaths of $[1,w]$} as the deaths which occurred after day $w$ but are attributed to infections in $[1,w]$.    This terminology is illustrated in Figure \ref{fig:intervals}.  We make the assumption that $\text{lag} < w$; thus residual deaths will not span multiple intervals. Therefore, the deaths subsequence, $(d_{w+1},d_{w+2},\dots,d_{2w})$, is the sum of the current deaths of $[w+1,2w]$ and the residual deaths of $[1,w]$.

We will compute the residual deaths, IFR, and lag for each interval from left to right.
To estimate the residual deaths of $[1,w]$, we first find the best-fit IFR and lag parameters $(a^*, b^*)$ over the interval  $[1,w]$, by calculating:
\begin{equation}\label{eq:firstintervalifr}
(a^*,b^*,\text{IFR}[1,w])=\bestfit((i_1,i_2,\dots,i_w),(d_1,d_2,\dots,d_w)),
\end{equation}
via Algorithm \ref{alg:rv_lag}. 
Using the best-fit IFR and lag, we calculate residual deaths of $[1,w]$ by time-shifting and vertically scaling the infections subsequence $(i_1,i_2,\dots,i_w)$ using a similar construction to $\shiftexpnovec{L}{\cdot}$ (see Equation \ref{eq:defn_shift_rv}), except that we allow the shifted vector to be longer than the initial vector.  For a random variable $L$ with maximum possible value $\ell_{\max}$, we will define the new elongated shift via 
$$\label{eq:longshift}\shiftexpext{L}{i_1,i_2,\dots,i_w}\equiv (i'_1,i'_2,\dots,i'_{w+\ell_{\max}}),$$
where each $i'_j$ is defined as in Equation \ref{eq:shift_rv_exp}. Thus to find the residual deaths of $[1,w]$, we compute $\text{IFR}[1,w]\cdot\shiftexpext{L}{i_1,i_2,\dots,i_w}$, where $L\sim$ Uniform($a^*,b^*)$, and take all entries after the $w^{th}$ entry.

Moving on to the next interval, we can now find the current deaths of $[w+1,2w]$ by subtracting the residual deaths of $[1,w]$ from the deaths subsequence $(d_{w+1},d_{w+2},\dots,d_{2w})$. We then repeat the process of computing residual deaths, IFR, and lag for intervals $[w+1,2w]$, $[2w+1,3w], \dots$ (see Algorithm \ref{alg:intervalsest}). 


\begin{algorithm}[H]
\caption{Computing the IFR in intervals of width $w$ using infections sequence $\vec{i}$ and deaths sequence $\vec{d}$ both of length $k$. We will use the notation $\vec{v}[s,t]$, for an arbitrary vector $\vec{v}$, to denote $(v_s,v_{s+1},v_{s+2},\dots,v_{t})$.}\label{alg:intervalsest}
\begin{algorithmic}
\STATE \texttt{curr\_l} $\leftarrow$ 0, \ \texttt{curr\_r} $\leftarrow$ w
\STATE initialize vector \texttt{residual\_deaths} $\leftarrow \vec{0}$
\WHILE{\texttt{curr\_r} $\le k$}
    \STATE initialize new vector $\vec{d'} \leftarrow \vec{d}[\texttt{curr\_l},\texttt{curr\_r}]$
    \FOR{$1 \le x \le \norm{\texttt{residual\_deaths}}$}
        \STATE $d'_x \leftarrow d'_x-\texttt{residual\_deaths}_x$
    \ENDFOR
    \STATE $(a^*,b^*,r^*) \leftarrow \bestfit(\vec{i}[\texttt{curr\_l},\texttt{curr\_r}],\vec{d'})$
    \STATE Output $\Uniform(a^*,b^*)$ and $r^*$ as the best-fit lag and IFR over $[\texttt{curr\_l},\texttt{curr\_r}]$
    \STATE Let $L$ denote a random variable distributed $\Uniform(a^*,b^*)$
    \STATE $\texttt{residual\_deaths} \leftarrow r^*\cdot\shiftexpext{L}{\vec{i}[\texttt{curr\_l},\texttt{curr\_r}]}[\texttt{curr\_r}+1,\texttt{curr\_r}+b^*]$
    \STATE $\texttt{curr\_l} \leftarrow \texttt{curr\_r}+1$, \
    $\texttt{curr\_r} \leftarrow \texttt{curr\_l}+w$
\ENDWHILE
\end{algorithmic}
\end{algorithm}

\section{Evaluation}
\label{sec:evaluation}

In this section, we apply our methodology for determining the IFR and the lag to data for several countries. We make use of 250 days of publicly available data on daily cases and deaths \cite{data-JHU} as well as on daily tests \cite{testing-change}. For each country, we will also make use of one antibody study.

\countryoverview{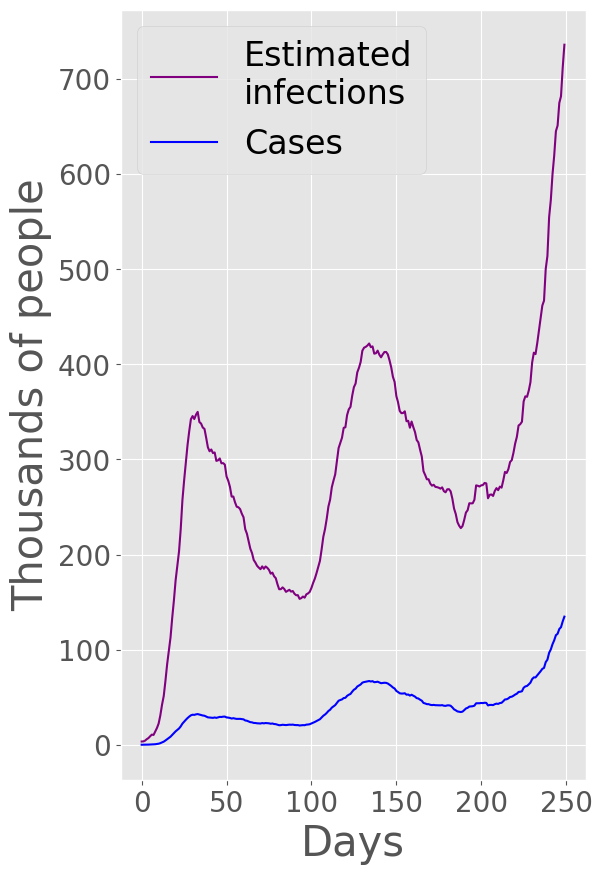}{USA}{the United States}{March 1, 2020}

\countryoverview{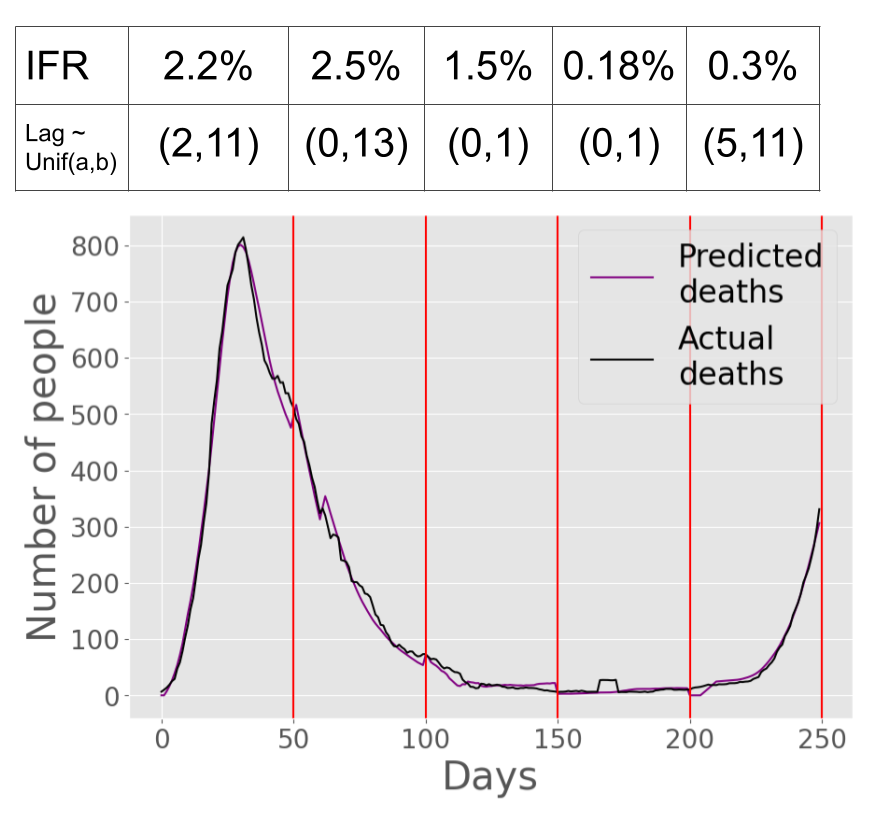}{Italy}{Italy}{March 1, 2020}

\countryoverview{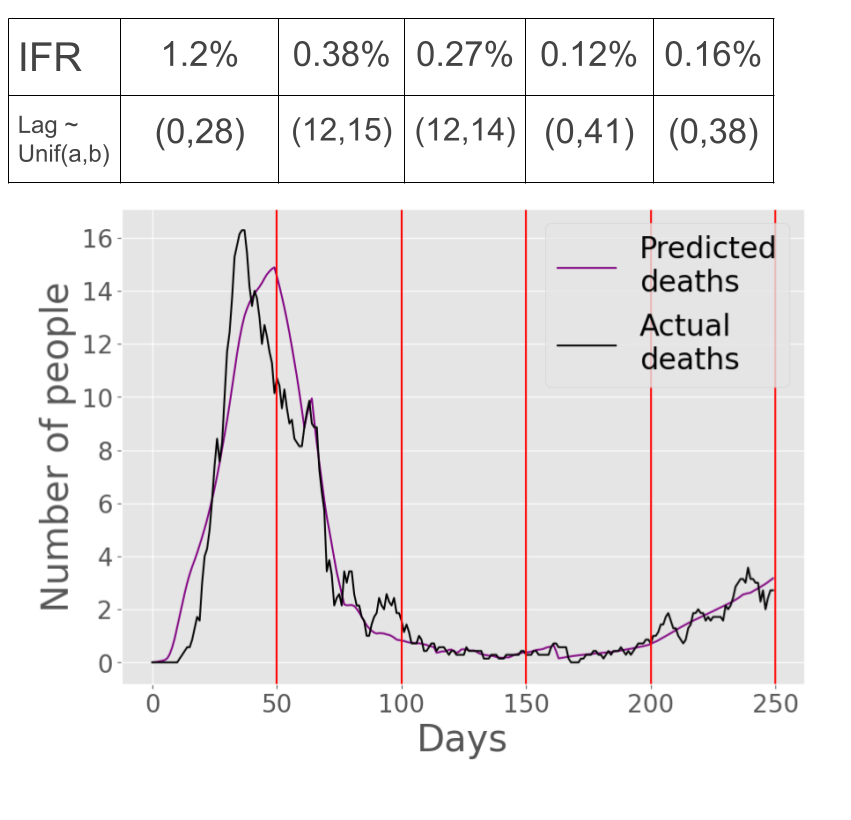}{Denmark}{Denmark}{March 1, 2020}
\countryoverview{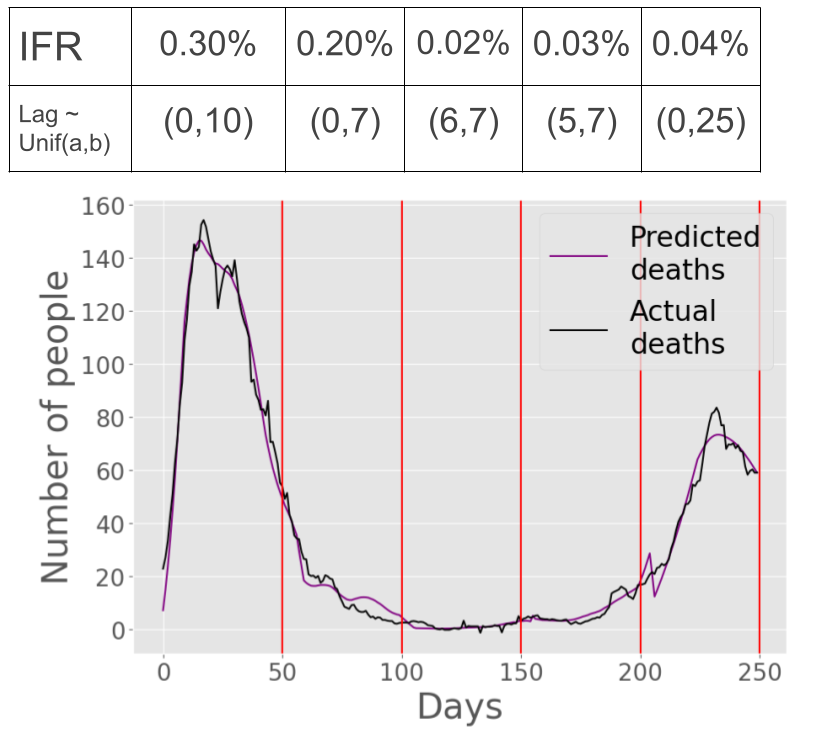}{Netherlands}{Netherlands}{March 22, 2020}

\textbf{United States} In calculating our estimated infections sequence, we assume a population of roughly 382 million \cite{worldometer}, 9\% of which had been infected by COVID before July 31, 2020 according to the prevalence of antibodies \cite{usa-antibody}.
In Figure \ref{fig:graph-USA}(a) we display the raw data.   The daily cases sequence, $\vec{c}$, is shown in blue and the daily deaths sequence, $\vec{d}$, is shown in black. In Figure \ref{fig:graph-USA}(b) we show the daily tests sequence, $\vec{t}$, in orange. Importantly, the number of daily tests increases a lot over time.   In Figure \ref{fig:graph-USA}(c) we show our estimated daily infections sequence, $\vec{i}$, in purple (as derived from Equation \ref{eq:calculateij}, where we computed $m=3.3$ to be optimal).   We juxtapose this with our daily cases sequence, $\vec{c}$, shown in blue. 
{\em Observe that the purple infection sequence is not simply a vertical scaling of the blue case sequence.} This is because we estimate the daily infections to be a function of {\em both} the daily cases and daily tests (see Section \ref{sec:inf-est}), and we can see that the daily tests increase steeply over time (see Figure \ref{fig:graph-USA}(b)).

In Figure \ref{fig:graph-USA}(d), we apply Algorithm \ref{alg:intervalsest} with an interval-width of $w = 50$ to our estimated infections sequence from Figure \ref{fig:graph-USA}(c) to obtain the  best-fit candidate deaths sequence (shown in purple).   We juxtapose this with the raw data for deaths, shown in black (same as what we saw in Figure \ref{fig:graph-USA}(a), but on different scale). The endpoints of each interval are shown as vertical red lines.  
Figure \ref{fig:graph-USA}(d) shows that the IFR decreased over time, eventually changing from 0.68\% to 0.24\%.   Furthermore, the lag is reasonably constant across intervals with the mean lag remaining around 8 days.
The best-fit candidate death sequence (purple line) is visually an excellent match to the actual deaths sequence (black line), indicating a low error in our model's optimization. The fact that the lag remains relatively constant across intervals is also an indication that we're not over-fitting.

\textbf{Italy}
We apply our same methodology to Italy (see
Figure \ref{fig:graph-Italy}). We assume a population of roughly 60 million \cite{worldometer}, 2.5\% of which was infected with COVID before June 20, 2020 according to the prevalence of antibodies \cite{italy-antibody}. As a result, we computed a value of $m=4.1$ to be optimal when applying Equation \ref{eq:calculateij}.
As in the United States, daily tests in Italy increase a lot over time. This causes the first peak in infections to make up a larger proportion of the total infections than the first peak in cases makes up in total cases.

Figure \ref{fig:graph-Italy}(d) shows that the IFR in Italy initially increased from 2.2\% to 2.5\% before dramatically dropping, even reaching as low as 0.18\%. The IFR estimates in Italy over the first three intervals are much higher than those of the U.S., which is consistent with news reports that Italian hospitals were overwhelmed during the beginning of the pandemic \cite{italy-hospitals}.  Note that the lag was fairly constant with a mean lag around 7 days, except for intervals 3 and 4 which had a lag of almost 0. The unrealistically small lag in these intervals can be attributed to over-fitting, since the deaths graph was essentially flat in these intervals. Time-shifting a flat line doesn't have much of an effect, so the algorithm selected a lag which was fairly arbitrary. Overall, however, the best-fit candidate death sequence (purple line) is visually an excellent match to the actual deaths sequence (black line), indicating that the estimates are generally accurate.

\textbf{Denmark} In Figure \ref{fig:graph-Denmark} we show our estimates for Denmark. We assume a population of roughly 5.8 million \cite{worldometer}, 1.1\% of which had been infected prior to May 15, 2020 according to the prevalence of antibodies \cite{denmark-antibodies}. As a result, we computed a value of $m=4.2$ to be optimal while applying Equation \ref{eq:calculateij}. Figure \ref{fig:graph-Denmark}(b) shows that the daily tests in Denmark rose steadily over time, having a similar effect on the infections sequence compared to the other countries. We estimate that the lag did not change much, with a mean of around 15 days across intervals. The IFR decreased over time changing from 1.2\% to 0.38\% to 0.27\% to 0.12\% to 0.16\%.

\textbf{Netherlands} In Figure \ref{fig:graph-Netherlands} we show our calculations for the Netherlands. We assume a population of roughly 17 million \cite{worldometer},  2.8\% of which was infected as of April 3, 2020 according to the prevalence of antibodies \cite{netherlands-antibodies}. As a result, we computed $m=2.2$ to be optimal in applying Equation \ref{eq:calculateij}. Importantly, we now let day 0 correspond to March 22, 2020 (rather than March 1), since this is when data on daily tests was first reported in the Netherlands \cite{testing-change}. Note that because a small amount of deaths could be attributed to infections that occurred before March 22, 2020, we likely overestimate the IFR in the first interval by a small amount and slightly underestimate the lag (since our model assumes all deaths in the first interval to be attributed to infections in the first interval). As in the other countries, daily tests trended upwards over time, however there was a slight decrease in daily tests towards the end of the pandemic (see Figure \ref{fig:graph-Netherlands}(b)). Figure \ref{fig:graph-Netherlands}(d) shows a visually excellent fit between the predicted deaths sequence and the actual deaths sequence, indicating that our model works well overall. We estimate that the lag stayed relatively constant (with mean of about $7$ days), except for a slight increase in the last interval. We estimate that the IFR decreased over time, changing from 0.30\% to 0.20\% to 0.02\% to 0.03\% to 0.04\%. These IFR estimates are much lower compared to other countries, but follow a similar trend in decrease over time.

\section{Conclusions and future work}\label{sec:conclusion}
Our approach aims to improve existing IFR estimates via several new ideas. We use data on the daily number of tests to help estimate infections. We treat the lag between cases and deaths as a random variable whose parameters we estimate empirically. We analyze small time intervals, allowing us to estimate a time-varying IFR and lag. In doing so, we find that the IFR in the U.S. decreases over time from 0.68\% to 0.24\%, with a mean lag between cases and deaths close to 8 days. Our model produces a visually excellent fit to the actual deaths sequence. We also achieve a good fit using data from Italy (IFR decreases from 2.2\% to 0.3\%), Denmark (IFR decreases from 1.2\% to 0.16\%), and the Netherlands (IFR decreases from 0.3\% to 0.04\%).

Overall, our approach for estimating IFR is very different from existing approaches and offers a number of new benefits. We are able to estimate IFR while relying on only a single antibody study, which can be performed at any time. Thus, we can consider an antibody study conducted early on in the pandemic before any individual's antibodies have faded. By contrast, existing approaches may rely on later antibody studies which underestimate the number of infections. Furthermore, our approach readily admits temporal changes. Existing approaches are unable to analyze smaller time-intervals. Lastly, our approach requires few data sources, and in particular doesn't require monitoring individuals. We rely on only a few key reports, each of which was easily accessible from the very start of the pandemic.

There are clearly opportunities for future work on this model. The approach introduced in this paper, for estimating the infections sequence, is novel, and we expect that others will build on this approach to make it more accurate.  There is also much work to be done on applying the approach to other countries and regions as more data becomes available.

\bibliographystyle{splncs04}

\input{main.bbl}
\end{document}

%% file: dependencies.tex
\usepackage[utf8]{inputenc}
\usepackage{float}
\usepackage{subfigure}
\usepackage[shortlabels]{enumitem}
\usepackage{amsmath, amsfonts, amssymb, graphicx}
\usepackage[subsection]{algorithm}
\usepackage{algorithmic}
\usepackage{multirow}
\usepackage{color}
\usepackage{url}
\usepackage{breakurl}
\usepackage[breaklinks]{hyperref}

%% file: macros-small.tex
\usepackage{url}

\usepackage{breakurl}
\usepackage[breaklinks]{hyperref}

\def\S{\mathcal{S}}

\def\l{\ell}

\providecommand{\norm}[1]{\left\Vert #1 \right\Vert}

\newcommand{\Uniform}{\mbox{Uniform}}

\DeclareMathOperator*{\argmin}{argmin}

\makeatletter
\newcommand\resetsubfigs{\setcounter{sub\@captype}{0}}
\makeatother

%% file: safe_macros.tex
\def\bestfit{\textsc{BestFit} }

\newcommand\shift[2]{\S_{#1}(\vec{#2})}
\newcommand\shiftexp[2]{\bar{\S}_{#1}(\vec{#2})}
\newcommand\shiftexpnovec[2]{\bar{\S}_{#1}(#2)}
\newcommand\shiftexpext[2]{\bar{\S}^{\text{long}}_{#1}(#2)}

\providecommand{\p}[1]{\textbf{P}\left\{ #1 \right\}}
\providecommand{\E}[1]{\textbf{E}\left[ #1 \right]}
\numberwithin{algorithm}{section}

%% file: macros-file-formats.tex
\newcommand{\evaluatedcountrypredictions}[3]{
    \begin{figure}
        \resetsubfigs
        \centering
        \subfigure[Cases (blue) and Estimated infections (purple)] {
            \includegraphics[height=0.18\paperheight]{segmented_complex_ifr/#1_inf_est.png}
        }
        \hspace{10mm}
        \subfigure[Predicted Deaths (purple) and Actual Deaths (black)]{
            \includegraphics[height=0.18\paperheight]{segmented_complex_ifr/#1_5_segmented.png}
        }
        \caption{#3} \label{#2}
    \end{figure}
}

\newcommand{\countryoverview}[4]{\begin{figure}
    \vspace*{-10pt}
    \resetsubfigs
    \centering
    \subfigure[Raw cases \& deaths]  {
        \includegraphics[width=0.18\textwidth]{raw_data/full_cfr_raw_#1.png}
    }\hspace{5mm}
    \subfigure[Daily tests]{\includegraphics[width=0.18\textwidth]{raw_data/tests_#1.png}}\\
    \subfigure[Estimated infections] {
        \includegraphics[width=0.18\textwidth]{segmented_complex_ifr/#1_inf_est.png}
    }\hspace{5mm}
    \subfigure[Candidate deaths vs. raw deaths sequence ] {
        \includegraphics[width=0.3\textwidth]{segmented_complex_ifr/#1_5_segmented.png}
    }
    \caption{Illustrating our algorithm for determining IFR and lag for #3 starting #4.}\label{fig:graph-#2}
    \vspace{-20pt}
\end{figure}}

%% file: main.bbl
\begin{thebibliography}{10}
\providecommand{\url}[1]{\texttt{#1}}
\providecommand{\urlprefix}{URL }
\providecommand{\doi}[1]{https://doi.org/#1}

\bibitem{antibodies-fade}
Clinical and immunological assessment of asymptomatic {SARS-CoV-2} infections.
  Nat Med  \textbf{26},  1200--1204

\bibitem{worldometer}
Countries in the world by population.
  \url{https://www.worldometers.info/world-population/population-by-country/}
  (November 2020)

\bibitem{italy-antibody}
Covid-19, illustrati i risultati dell'indagine di sieroprevalenza.
  \url{salute.gov.it/portale/nuovocoronavirus/dettaglioNotizieNuovoCoronavirus.jsp?id=4998}
  (2020)

\bibitem{death-rate-change}
Data show hospitalized {Covid-19} patients are surviving at higher rates, but
  surge in cases could roll back gains.
  \url{https://www.statnews.com/2020/11/23/hospitalized-covid-19-patients-surviving-at-higher-rates-but-surge-could-roll-back-gains/}
  (November 2020)

\bibitem{why-important}
Estimating mortality from {COVID-19}.
  \url{https://www.who.int/news-room/commentaries/detail/estimating-mortality-from-covid-19}
  (August 2020)

\bibitem{italy-hospitals}
Italy's hospitals overwhelmed by coronavirus as death toll soars.
  \url{https://www.cbsnews.com/video/italys-hospitals-overwhelmed-by-coronavirus-as-death-toll-soars/}
  (2020)

\bibitem{owid}
Our world in data coronavirus pandemic data explorer.
  \url{https://ourworldindata.org/coronavirus-data-explorer} (November 2020)

\bibitem{false-pos}
Reasons for a false positive or false negative {COVID-19} test result.
  \url{https://www.uchealth.com/en/media-room/videos/reasons-for-a-false-positive-or-false-negative-covid-19-test-result}
  (August 2020)

\bibitem{who-report}
Report of the {WHO-China} joint mission on coronavirus disease 2019
  ({COVID-19}).
  \url{https://www.who.int/publications/i/item/report-of-the-who-china-joint-mission-on-coronavirus-disease-2019-(covid-19)}
  (February 2020)

\bibitem{states-slow}
Daily updates of totals by week and state.
  \url{https://www.cdc.gov/nchs/nvss/vsrr/covid19/index.htm} (February 2021)

\bibitem{usa-antibody}
Anand, S., Montez-Rath, M., Han, J., Bozeman, J., Kerschmann, R., Beyer, P.,
  Parsonnet, J., Chertow, G.M.: Prevalence of {SARS-CoV-2} antibodies in a
  large nationwide sample of patients on dialysis in the {USA}: a
  cross-sectional study. Lancet Inf Dis.  \textbf{369},  1335--1344 (2020)

\bibitem{data-JHU}
Dong, E., Du, H., Gardner, L.: An interactive web-based dashboard to track
  {COVID-19} in real time. Lancet Inf Dis.  \textbf{20}(5),  533--534 (2020)

\bibitem{denmark-antibodies}
Espenhain, L., Tribler, S., Jorgensen, C.S., Holm~Hansen, C., Wolff~Sonksen,
  U., Ethelberg, S.: Prevalence of {SARS-CoV-2} antibodies in {Denmark} 2020:
  results from nationwide, population-based sero-epidemiological surveys.
  \url{https://www.medrxiv.org/content/10.1101/2021.04.07.21254703v1}

\bibitem{testing-change}
Hasell, J., Mathieu, E., Beltekian, D., Macdonald, B., Giattino, C.,
  Ortiz-Ospina, E., Roser, M., Ritchie, H.: A cross-country database of
  {COVID-19} testing. Scientific Data  \textbf{7}(345) (2020)

\bibitem{Levin}
Levin, A.T., Hanage, W.P., Owusu-Boaitey, N., Cochran, K.B., Walsh, S.P.,
  Meyerowitz-Katz, G.: Assessing the age specificity of infection fatality
  rates for {COVID-19}: Systematic review, meta-analysis, and public policy
  implications.
  \url{https://www.medrxiv.org/content/10.1101/2020.07.23.20160895v7} (2020)

\bibitem{Marra}
Marra, V., Quartin, M.: A {Bayesian} estimate of the {COVID-19} infection
  fatality rate in {Brazil} based on a random seroprevalence survey. medRxiv
  (2020). \doi{10.1101/2020.08.18.20177626},
  \url{https://www.medrxiv.org/content/early/2020/10/09/2020.08.18.20177626}

\bibitem{MeyerowitzKatz}
Meyerowitz-Katz, G., Merone, L.: A systematic review and meta-analysis of
  published research data on {COVID-19} infection fatality rates. Int J Infect
  Dis  \textbf{Dec.}(101),  138--148 (2020)

\bibitem{Newall}
Newall, A., Leong, R., Nazareno, A., Muscatello, D., Wood, J., Kim, W.:
  Estimating the infection and case fatality ratio for coronavirus disease
  ({COVID-19}) using age-adjusted data from the outbreak on the {Diamond
  Princess} cruise ship, {February} 2020. Int J Infect Dis  \textbf{Dec.}(101),
   306--311 (2020)

\bibitem{not-randomized}
Padula, W.V.: Why only test symptomatic patients? {Consider} random screening
  for {COVID-19}. Appl Health Econ Health Policy  \textbf{18}(3),  333--334
  (2020)

\bibitem{deaths-unattributed}
Rossen, L.M., Branum, A.M., Ahmad, F.B., Sutton, P., Anderson, R.N.: Excess
  deaths associated with {COVID-19}, by age and race and ethnicity — {United
  States}, {January 26–October 3}, 2020. Morbidity and Mortality Weekly
  Report  \textbf{69}(42),  1522--1527 (2020)

\bibitem{Russell}
Russell, T.W., Hellewell, J., Jarvis, C.I., van Zandvoort, K., Abbott, S.,
  Ratnayake, R., working group, C.C.., Flasche, S., Eggo, R.M., Edmunds, W.J.,
  Kucharski1, A.J.: Delay-adjusted age- and sex-specific case fatality rates
  for {COVID-19} in {South Korea}: Evolution in the estimated risk of mortality
  throughout the epidemic. Euro Surveill  \textbf{25}(12) (2020)

\bibitem{Villa}
Villa, M., Myers, J.F., Turkheimer, F.: {COVID-19}: Recovering estimates of the
  infected fatality rate during an ongoing pandemic through partial data.
  \url{medrxiv.org/content/10.1101/2020.04.10.20060764v1} (2020)

\bibitem{netherlands-antibodies}
Vos, E.R.A., den Hartog, G., Schepp, R.M., Kaaijk, P., van Vliet, J., Helm, K.,
  Smits, G., Wijmenga-Monsuur, A., Verberk, J.D.M., van Boven, M., van
  Binnendijk, R.S., de~Melker, H.E., Mollema, L., van~der Klis, F.R.M.:
  Nationwide seroprevalence of {SARS-CoV-2} and identification of risk factors
  in the general population of the {Netherlands} during the first epidemic
  wave. Journal of Epidemiology \& Community Health  \textbf{75}(6),  489--495
  (2021). \doi{10.1136/jech-2020-215678},
  \url{https://jech.bmj.com/content/75/6/489}

\end{thebibliography}
